# Deutsch's algorithm with topological charges of optical vortices via non-degenerate four-wave mixing

Mingtao Cao,[1] Liang Han,[1] Ruifeng Liu,[1] Hao Liu,[1] Dong Wei,[1] Pei Zhang,[1,4] Yu Zhou,[1] Wenge Guo,[2] Shougang Zhang,[3] Hong Gao,[1,5] and Fuli Li[1]

[1]*Department of Applied Physics, Xi'an Jiaotong University, Xi'an 710049, China*
[2]*MOE Key Laboratory for Electricity Gas and Oil Logging, Xi'an Shiyou University, Xi'an 710065, China*
[3]*CAS Key Lab Time & Frequency Primary Standard, National Time Service Center, Xi'an 710600, China*
[4]*zhangpei@mail.ustc.edu.cn*
[5]*honggao@mail.xjtu.edu.cn*

**Abstract:** We propose a scheme to implement the Deutsch's algorithm through non-degenerate four-wave mixing process. By employing photon topological charges of optical vortices, we demonstrate the ability to realize the necessary four logic gates for all balanced and constant functions. We also analyze the feasibility of the proposed scheme on the single photon level.

**OCIS codes:** (270.5585) Quantum information and processing; (190.4380) Nonlinear optics, four-wave mixing.

## 1. Introduction

Light with a Laguerre-Gaussian (LG) mode carries orbital angular momentum (OAM) [1]. The OAM of light can be identified by a series of integer quantum number, $l$, which form a complete basis set in Hilbert space. This unique property has shown immense potential applications in quantum information and quantum computation [2, 3]. So far, OAM entangled states have been realized through a spontaneous parametric down conversion (SPDC) process [4], and OAMs of generated photon pairs also have been proven to be correlated through the four-wave mixing (FWM) process in atomic ensembles [5]. Due to the narrow bandwidth of the generated OAM photons, the nonlinear process in an atomic ensemble has attracted great attention in recent years. Moreover, as an essential request in quantum computation, OAM transferring between atoms and photons has also been explored [5-9]. The topological charges of optical vortices (i.e., quantum number $l$ of OAM) have been proven to obey OAM conservation during the transferring process [10, 11]. Based on above transformations, the computation of topological charges of two optical vortices has been achieved via a non-degenerate FWM process in reference [12], which presents some potential applications in quantum computing, such as quantum Deutsch's algorithm [13]. In previous realizations of quantum Deutsch's algorithm, many physical systems have been exploited, including ion traps [14], semiconductor quantum dots [15], cold atoms [16], and linear optics [17-19]. However, the qubit states in those systems are not readily accessed through an atomic medium, which is a key point for quantum repeater and long distance quantum communication.

In this paper, we present an experimental scheme to realize the logical operations of Deutsch's algorithm in atomic ensembles through a non-degenerate FWM process. We demonstrate the realization of the four logic gates used for implementing the Deutsch's algorithm. Since we employ the OAM as the qubit, the results can be easily detected by viewing the output image. On the single photon level, the state can be detected by using the state projection method. The scheme we proposed may provide a useful method for quantum computation with a multidimensional system.

## 2. Experimental scheme

Our experimental proposal to realize the Deutsch's algorithm utilizes a non-degenerate FWM process. A typical FWM process is schematically shown in Fig.1(a). Two pump beams (P1 and P2) with orthogonal polarizations counter-propagate along the axis of the atomic cell. The probe beam (P3) with the same polarization as P1 joins the cell at a small angle with pump fields. According to the phase matching condition, FWM signal (S) that counter-propagates with P3 can be generated. The phase matching condition also requires that the polarization of S field is naturally orthogonal with P3 field. Then we can place a polarizing beam-splitter

(PBS) in the path of S field to split the S and P3, as shown in Fig. 1(a). The energy diagram of atom levels coupled by different laser fields in the scheme is shown in Fig. 1(b). In FWM process nonlinear susceptibility greatly relies on the first order of coherent term [20]. Thus it is essential to hold the quantum coherence in non-degenerated FWM process [12]. This will also benefit the following quantum algorithm.

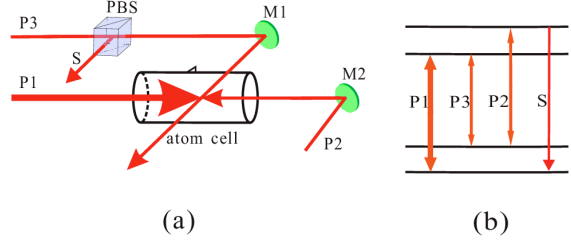

Fig.1. (a) The experimental setup of the four-wave mixing scheme. P1, P2, P3 field are provided by three lasers. M1, M2 are mirrors. PBS is polarizing beam-splitter, which transmits horizontal polarization and reflects vertical polarization. (b) The energy diagram of atom levels coupled by different laser fields in the scheme.

As we know, OAM states can be generated by varying methods, such as computer-generated hologram (CGH), spiral phase plate, Q-plate, etc [21]. For CGH, the first order of diffracted photon can obtain OAM of $+h$ or $-h$. The diffracted field carrying OAMs can be expressed by [21]:

$$LG_p^l = \sqrt{\frac{2p!}{\pi(p+|l|)!}} \frac{1}{w(z)} \left[\frac{r\sqrt{2}}{w(z)}\right]^{|l|} \exp\left[\frac{-r^2}{w^2(z)}\right] L_p^{|l|}(\frac{2r^2}{w^2(z)}) \\ \exp\left[\frac{ik_0 r^2 z}{2(z^2+z_R^2)}\right] \exp\left[-i(2p+|l|+1)\tan^{-1}(\frac{z}{z_R})\right] \exp[il\phi], \tag{1}$$

where $w(z) = w(0)[(z^2+z_R^2)/z_R^2]^{1/2}$ is the beam spot size, $z_R$ is Rayleigh range, and $(2p+|l|+1)\tan^{-1}(Z/Z_R)$ denotes Gouy phase. Associated Laguerre polynomial $L_p^{|l|}(x)$ is obtained by $L_p^{|l|}(x) = (-1)^{|l|}\frac{d^{|l|}}{dx^{|l|}}L_{p+|l|}(x)$ and $L_{p+|l|}(x)$ is the Laguerre polynomial. $l$ is the azimuthal index giving an OAM of $l h$ per photon, and $p+1$ is the number of radial nodes in the intensity distribution. For simplicity and without loss of generality, we set $p = 0$ all through the manuscript, and only OAM number $l$ is considered. In our following scheme we mainly concern the OAM exchange in non-degenerate FWM process, thus only the transverse phase term $\exp[il\phi]$ of laser fields is taken into account. Therefore, P1, P2, and P3 fields with different OAMs can be written as:

$$E_{P1} = A_1 e^{-il_1\phi}, \quad E_{P2} = A_2 e^{-il_2\phi}, \quad E_{P3} = A_3 e^{-il_3\phi}, \tag{2}$$

where $A_i = \sqrt{\frac{2}{\pi|l_i|!}}\frac{1}{w(z)}\left[\frac{r\sqrt{2}}{w(z)}\right]^{|l_i|}\exp\left[\frac{-r^2}{w^2(z)}\right]L_0^{|l_i|}(\frac{2r^2}{w^2(z)})\exp\left[\frac{ik_0 r^2 z}{2(z^2+z_R^2)}\right]\exp\left[-i(|l_i|+1)\tan^{-1}(\frac{z}{z_R})\right]$, for $i = 1, 2, 3$. Since $A_i$ only contributes to the amplitude of the field, it can be departed from the transverse phase. The output signal field is determined by both the third-order nonlinearity $\chi^{(3)}$ and the input field. We note that the nonlinearity $\chi^{(3)}$ can be effectively enhanced at the cost of greatly reducing $\chi^{(1)}$ due to the electromagnetically induced transparency (EIT) effect [22]. Thus it offers us a great advantage to obtain higher transform efficiency during the FWM process. The FWM signal field follows [23]:

$$E_S = \chi^{(3)} E_{P1} E_{P2} E_{P3}^*. \tag{3}$$

Now we put the transverse phase of each field into Eq. (3), the generated signal field $E_S$ can be obtained as:

$$E_S = \chi^{(3)} A_1 \cdot A_2 \cdot A_3^* \exp[-i(l_1 + l_2 - l_3)\phi]. \quad (4)$$

Thus, the OAM transferring in the process can be described as: $l_S = l_{p1} + l_{p2} - l_{p3}$, which is the fundamental relationship to realize different quantum gates in our scheme.

Tab.1 Encoding protocol using quantum states of OAM to realize the Deutsch's algorithm.

| Class | $l$ | Spatial mode | Computational basis |
|---|---|---|---|
| Controlling bit (P3) | 0 | \|  ⟩ | \|0⟩ |
|  | 1 | \|  ⟩ | \|1⟩ |
| Target bit (P2,S) | 0 | \|  ⟩ | \|0⟩ |
|  | $1 \, or \, -1$ | \|  ⟩ | \|1⟩ |

## 3. Realization of quantum Deutsch's algorithm

### 3.1. Coding rule for the logic gates and theoretical analysis

To realize Deutsch's algorithm in a non-degenerate FWM system, we choose different OAMs to encode the qubit states. The encoding protocol of control and target qubits is shown in Tab. 1. We choose photon's OAM of probe field (P3) as the control qubit, where the photon carrying OAMs of 0ħ is defined as $|0\rangle$, and photon carrying +ħ is $|1\rangle$. The photon's OAM of backward field (P2) and signal field (S) are selected as the target qubits, where the photon carrying OAM of 0ħ is defined as $|0\rangle$, and photon carrying ±ħ is $|1\rangle$.

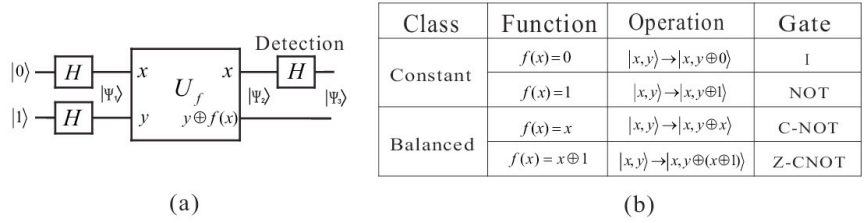

(a)                                                              (b)

Fig.2. (a) Quantum circuit for Deutsch's algorithm. $H$ is the Hadamard gate to supply the superposition state, $U_f$ is quantum operation, which takes inputs $|x,y\rangle$ to $|x, y \oplus f(x)\rangle$ for the four possible functions and output states $|\psi_2\rangle$. $|\psi_1\rangle$ is the converted state after Hadamard gate, and $|\psi_3\rangle$ is the final state to be measured. (b) Four basic logical operations for Deutsch's algorithm.

Deutsch's algorithm can be thought as a solution to combine quantum parallelism with a property of quantum interference. Figure 2(a) is the quantum circuit for Deutsch's algorithm. When the input state is $|\psi_0\rangle = |0\rangle|1\rangle$, the final transformed state is $|\psi_3\rangle = \pm|f(0) \oplus f(1)\rangle(|0\rangle - |1\rangle)/\sqrt{2}$, where '⊕' represents Boolean addition and $f(x)$ is a Boolean function. To realize the algorithm, we need a setup to implement the $U_f$ operations for the four possible $f(x)$ functions, which are shown in Fig. 2(b). The four possible actions of $f(x)$ can be classified as two types: constant functions and balanced functions. For a constant function, $f(x)$ is constant, such as $f(x) = 0$ or $f(x) = 1$ regardless of the input states. While the function would be balanced if the operation agrees with $f(x) = x$ or $f(x) = x \oplus 1$. Clearly, above four functions correspond to Identity (I), NOT, Controlled-NOT (C-NOT) and Zero-controlled NOT (Z-CONT) gates, respectively. The goal of Deutsch's algorithm is to find whether the function is constant or balanced. The quantum computer determines $f(x)$ to be balanced or constant by measuring the first qubit of $|\psi_3\rangle$ only once against the twice evaluating of $f(x)$ for classical computation. To implement the Deutsch's algorithm, we need to realize the four logic gates in advance.

*3.2. The experimental scheme to realize the four logic gates*

Our experimental scheme is shown in Fig.3. It contains three parts: initial states preparation, operations and detection. We can use a special displaced CGH (DH) as the Hadamard gate, after which we can get state $|\psi_1\rangle=(|0\rangle+|1\rangle)\cdot(|0\rangle-|1\rangle)/\sqrt{2}$. Then $|\psi_1\rangle$ will be transformed by $U_f$ operations for the four possible $f(x)$ functions. During the detecting process, we can use a single mode fiber combining with a CGH to collect the control bit photons. The implementations of the four operations are shown in Fig.4 in detail.

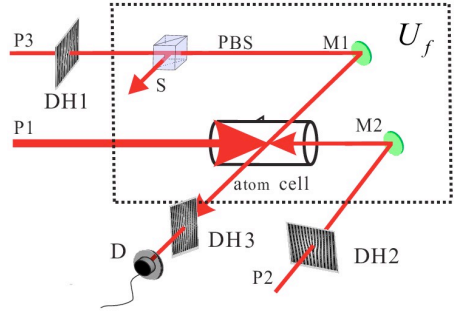

Fig.3. Experimental scheme of Deutsch's algorithm. DH1, DH2, and DH3 are three displaced holograms, and they are chosen as the Hadamard gate to supply the superposition state. Dashed square area is the experimental implementation of $U_f$ operations, in which four logic gates can be realized in different methods. D represents the photon detector.

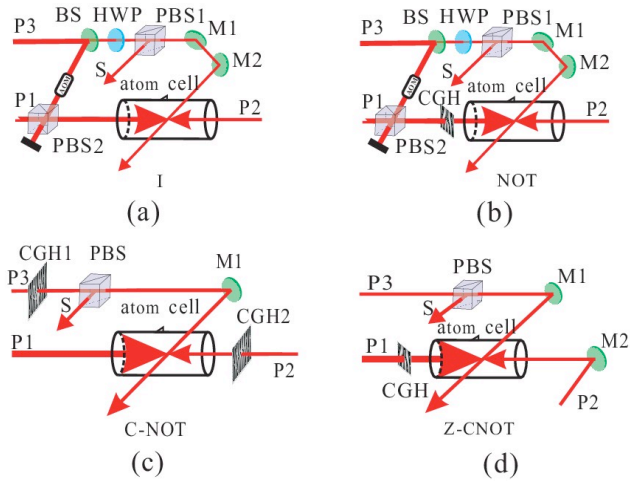

Fig.4. Experimental realization of $U_f$ in a FWM process, (a) I gate, (b) NOT gate, (c) C-NOT gate, and (d) Z-CNOT gate. CGH is computer-generated hologram, which is applied to change the photon's OAM. In the I gate and the NOT gate, an acoustic-optical modulator (AOM) can be introduced to make the frequency shift, so the P1 and P3 fields can be manipulated simultaneously. In the I and the NOT gates, half wave plate (HWP) is placed behind the BS to change the P3 field's polarization. Mirrors in both gates are introduced to change the sign of OAM. In the C-NOT gate, CGH1 and CGH2 are introduced to prepare the arbitrary initial sates in OAM space.

The identity (I) gate means that target qubit is never changed. The OAM transition relation $l_S = l_{p1} + l_{p2} - l_{p3}$ can be reduced to $l_S = l_{p2}$, if the OAM of P1 and P3 fields are set identically. Thus the OAM of S and P2 fields can hold the same value, which means that the generated photon state maintains the target state. The way to implement the I gate is shown in Fig. 4(a). P1 and P3 fields are separated from a same seed beam by a high reflection beam splitter (BS), thus the photon's OAM of P1 and P3 fields can be manipulated simultaneously. Since the scheme is based on non-degenerate FWM, P1 and P3 need to couple to different levels. This can be achieved by introducing an acoustic-optical modulator (AOM) to shift the frequency, so that P1 field can couple to the corresponding transition. In an experimental realization, the polarization of P3 would be set to vertical, and would pass through PBS1 by adjusting the half wave plate (HWP). Moreover, the polarization of P1 would also be set to vertical, so it would be reflected by PBS2 without joining in the mixing process. Therefore, no matter what the control qubit state is, the generated photon state maintains the same state as target.

$$|0,0\rangle = |\ ,\ \rangle \xrightarrow{I} |\ ,\ \rangle = |0,0\rangle,\quad |1,0\rangle = |\ ,\ \rangle \xrightarrow{I} |\ ,\ \rangle = |1,0\rangle,$$
$$|0,1\rangle = |\ ,\ \rangle \xrightarrow{I} |\ ,\ \rangle = |0,1\rangle,\quad |1,1\rangle = |\ ,\ \rangle \xrightarrow{I} |\ ,\ \rangle = |1,1\rangle. \quad (5)$$

The NOT gate means that generated signal photon acquires the opposite state against the state of target qubit. In order to realize the quantum NOT gate, we set $l_{p1} = l_{p3} - 1$. So the OAM transition relation can be reduced to $l_S = l_{p2} - 1$. The experimental realization is shown in Fig. 4(b). The NOT gate and the I gate are opposite operations, thus the way to realize them is similar. However, we need to add an additional CGH in the path of P1 to increase the OAM by $h$. In such way, we find that no matter what the control qubit state is, the generated signal photon acquires the opposite state against the state of target qubit. Thus the NOT gate also works.

$$|0,0\rangle = |\ ,\ \rangle \xrightarrow{NOT} |\ ,\ \rangle = |0,1\rangle,\quad |1,0\rangle = |\ ,\ \rangle \xrightarrow{NOT} |\ ,\ \rangle = |1,1\rangle,$$
$$|0,1\rangle = |\ ,\ \rangle \xrightarrow{NOT} |\ ,\ \rangle = |0,0\rangle,\quad |1,1\rangle = |\ ,\ \rangle \xrightarrow{NOT} |\ ,\ \rangle = |1,0\rangle. \quad (6)$$

The realization of the controlled-NOT (C-NOT) gate is shown in Fig. 4(c). We set the OAM of P1 field to $0h$, so the OAM transition relation is $l_S = l_{p2} - l_{p3}$. This shows that the target qubit flips when the control qubit is logical value '1', and maintains when the control qubit is logical value '0'. In the setup, we can also insert a CGH1 in the path of P3 beam to reduce the OAM by $h$, and a CGH2 in the path of P2 beam to add the OAM by $h$. Considering the reflective mirror effect, after the photon of P3 field carrying OAM of $0h$ ($h$) passes through CGH1 and reflected by M1, its OAM contributing to the FWM process is actually changed to $h$ ($0h$). Therefore, when the control qubit is $|0\rangle$, it can be easily obtained from Eq. (4) that the generated signal photon maintains the same state as the target qubit. On the other hand, when the control qubit is $|1\rangle$, the generated signal photon acquires the opposite state against the state of the target qubit. Noting that OAM of $-h$ is also assigned as $|1\rangle$ state according to the encoding rule. In brief, the above operation follows the C-NOT logic transformation law as shown in Fig. 2(b). To show how the quantum C-NOT gate works, we input an initial state as $\alpha|0\rangle + \beta|1\rangle$ for the target photon, where $\alpha^2 + \beta^2 = 1$. Then the generated photon state can be obtained as:

$$\hat{P}[\alpha|0\rangle + \beta|1\rangle] = \alpha|e^{-i(0+1-1)\phi}\rangle + \beta|e^{-i(0+0-1)\phi}\rangle = \alpha|0\rangle + \beta|1\rangle,$$
$$\hat{P}'[\alpha|0\rangle + \beta|1\rangle] = \alpha|e^{-i(0+1-0)\phi}\rangle + \beta|e^{-i(0+0-0)\phi}\rangle = \alpha|1\rangle + \beta|0\rangle. \quad (7)$$

Here, the operator $\hat{P}$ ($\hat{P}'$) denotes the FWM operation when the initial control qubit is $|0\rangle$

($|1\rangle$). The results confirm that the quantum C-NOT logic transformation works as follow.

$$|0,0\rangle=|\quad,\quad\rangle\xrightarrow{C-NOT}|\quad,\quad\rangle=|0,0\rangle,\quad |1,0\rangle=|\quad,\quad\rangle\xrightarrow{C-NOT}|\quad,\quad\rangle=|1,1\rangle,$$
$$|0,1\rangle=|\quad,\quad\rangle\xrightarrow{C-NOT}|\quad,\quad\rangle=|0,1\rangle,\quad |1,1\rangle=|\quad,\quad\rangle\xrightarrow{C-NOT}|\quad,\quad\rangle=|1,0\rangle. \quad (8)$$

Figure 4(d) shows the way to achieve the Z-CNOT gate which is short for zero-controlled NOT gate. The OAM transition relation of the Z-CNOT gate is $l_S = l_{p2} - l_{p3} + 1$, if we set the OAM of P1 field to h. It is clear that the target qubit flips when the control qubit is logical value '0', and maintains when the control qubit is logical value '1'. In the setup, a mirror M2 is introduced to change the sign of OAM of P2 field. Similarly, after a photon from the P2 field, initially carrying OAM of 0h ( h ), is reflected by M2 and passes through CGH, its OAM contributing to the FWM process is 0h ( –h ). We can prove that the Z-CONT gate works from similar derivation process of the C-NOT gate.

$$|0,0\rangle=|\quad,\quad\rangle\xrightarrow{Z-CNOT}|\quad,\quad\rangle=|0,1\rangle,\quad |1,0\rangle=|\quad,\quad\rangle\xrightarrow{Z-CNOT}|\quad,\quad\rangle=|1,0\rangle,$$
$$|0,1\rangle=|\quad,\quad\rangle\xrightarrow{Z-CNOT}|\quad,\quad\rangle=|0,0\rangle,\quad |1,1\rangle=|\quad,\quad\rangle\xrightarrow{Z-CNOT}|\quad,\quad\rangle=|1,1\rangle. \quad (9)$$

## 4. Result testing

Suppose the input state is $|0\rangle \cdot |1\rangle$, after the four operations, the output states $|\psi_2\rangle$ and $|\psi_3\rangle$ can be expressed as:

$$|\psi_2\rangle = \begin{cases} \frac{1}{2}(|0\rangle+|1\rangle)\cdot(|0\rangle-|1\rangle) \\ -\frac{1}{2}(|0\rangle+|1\rangle)\cdot(|0\rangle-|1\rangle) \\ \frac{1}{2}(|0\rangle-|1\rangle)\cdot(|0\rangle-|1\rangle) \\ -\frac{1}{2}(|0\rangle-|1\rangle)\cdot(|0\rangle-|1\rangle) \end{cases} \xrightarrow{H} |\psi_3\rangle = \begin{cases} |0\rangle\cdot\frac{(|0\rangle-|1\rangle)}{\sqrt{2}} & \text{I gate,} \\ -|0\rangle\cdot\frac{(|0\rangle-|1\rangle)}{\sqrt{2}} & \text{NOT gate,} \\ |1\rangle\cdot\frac{(|0\rangle-|1\rangle)}{\sqrt{2}} & \text{C-NOT gate,} \\ -|1\rangle\cdot\frac{(|0\rangle-|1\rangle)}{\sqrt{2}} & \text{Z-CNOT gate.} \end{cases} \quad (10)$$

We can clearly see that the first qubit of the output states are orthogonal for the two different classes of $f(x)$. So we define the first orthogonal qubit of output states as the testing state. If we detect $|\psi_2\rangle$, for the I gate and the NOT gate operations, the testing state is $|0\rangle+|1\rangle$, which can be obtained from $|\psi_2\rangle$ in Eq. (10). That is to say, we can observe the interferential mode of 0h and h OAM photons in constant function case. On the other hand, for the C-NOT gate and the Z-CNOT gate operations, the testing state is $|0\rangle-|1\rangle$. In this balanced function case the control qubit state is presented as the interferential mode of 0h and -h OAM photons. Figure 5(a) is the mode pattern of testing states for the I gate and the NOT gate, which presents as the result of constant function. Figure 5(b) is the mode pattern of testing states for the C-NOT gate and the Z-CNOT gate, which can be labeled as the result of balanced function. Final testing states of $|\psi_3\rangle$ in Eq. (10) are shown in Fig. 5(c) and (d). Figure 5(c) is the mode pattern of testing states for constant function, which presents as Gaussian distribution mode. Figure 5(d) is the mode pattern of testing states for balanced function, which presents as doughnut distribution mode. On the single photon level, we can measure the testing states by single photon projection setup, which is comprised of a CGH and a single mode fiber with photon detector [4, 24]. In this way, we can achieve our goal to distinguish the two kinds of functions (constant and balanced) by one step of calculation.

According to the above realization of the four logic gates, we can implement the Deutsch's algorithm by using the single photon's OAM state. On the single photon level, there are two significant aspects that should be considered: success probability of the logic gates and the additional noise caused by spontaneous radiation.

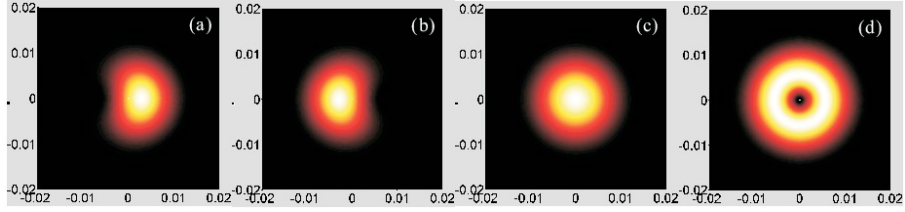

Fig.5. Testing results of two types of function. (a) and (b) are the intensity distributions of constant function and balance function when we detect the testing states of $|\psi_2\rangle$, respectively. While (c) and (d) are intensity distributions of constant function and balanced function when we detect the testing states of $|\psi_3\rangle$.

The success probability of the logic gates mainly relies on the relative transition ratio in FWM process, which is restricted by the strength of the nonlinearity, the strong pump fields, etc. Ideally, the relative transition ratio can be calculated from the dipole matrix coupled with different atomic hyperfine levels. For instance, when considering D1 line of $^{85}$Rb as the energy diagram in Fig. 1(b), the calculated relative transition ratio for signal field is 83.1% [25]. However, this number would be correspondingly reduced in experimental realization due to the nonlinearity and the pump noise. And the success probability of the gates will not be as high as the calculated relative transition ratio, but it is still operational. For example, a more than 60% of the strength of the nonlinearity has been observed in relative strong pump field experiment, in which $^{85}$Rb chose as the medium to interact with light [20]. Considering the noise on the single photon level, there are two methods to eliminate them. Firstly, the noise photons will carry different OAM from signal photons. They can be filtered out in projection detection. Secondly, the frequency of noise photons is far away from the signal photons (in $^{85}$Rb system the frequency difference is around GHz), thus we can filter the noise photons with a multi-pass Fabry-Pérot etalon. It has been demonstrated that the extremely large total suppression of the pump beam can reach 118 dB signal-to-noise ratio [26]. Moreover, the fidelity of reconstructed entangled OAM states has been proven to be more than 80% in spontaneous four-wave mixing system [5]. So we believe our scheme is feasible on the single photon level.

## 5. Conclusion

We propose an experimental scheme to realize the Deutsch's algorithm through a non-degenerate four-wave mixing system. In the nonlinear medium, the OAM of photons can be transferred following the conservation law, which shows the potential applications in quantum computing. We demonstrate theoretically the ability to realize the four logic gates used to implement the Deutsch's algorithm in atomic ensembles. By detecting the mode patterns of the testing states, we divide the four logic gates into two kinds of function (constant and balanced), in one step. We also investigate the feasibility of the proposed scheme on the single photon level in $^{85}$Rb vapor. Our scheme is an important element in quantum networking with atomic ensembles, and we believe that the proposal can also be applied to quantum computation with higher-dimensional quantum system using OAM of photons.

**ACKNOWLEDGMENTS**

We acknowledge financial support from the National Natural Science Foundation of China (NSFC) under grants 11074198 and 11004158, key program of NSFC under grant 10834007, NSFC for Distinguished Young Scholars of China under grant 61025023, and Special Prophase Project on the National Basic Research Program of China under grant 2011CB311807.